\documentclass[baaa]{baaa}

\usepackage[pdftex]{hyperref}
\usepackage{subfigure}
\usepackage{natbib}
\usepackage{helvet,soul}
\usepackage[font=small]{caption}

\contriblanguage{1}

\contribtype{1}

\thematicarea{1}

\title{Statistical analysis of global properties of galaxies in the direction of the Fornax cluster with S-PLUS}

\titlerunning{Global properties of galaxies in the direction of Fornax with S-PLUS}

\author{
J.P. Calderón\inst{1,2}, 
A.V. Smith Castelli\inst{1,2},
E.V.R. de Lima\inst{3},
A.R. Lopes\inst{4},
\linebreak
F. Almeida-Fernandes\inst{3,5}
\& 
C. Mendes de Oliveira\inst{3}
}

\authorrunning{Calderón et al.}

\contact{jpcalderon@fcaglp.unlp.edu.ar}

\institute{
Instituto de Astrof{\'\i}sica de La Plata, CONICET--UNLP, Argentina
\and
Facultad de Ciencias Astron\'omicas y Geof{\'\i}sicas, UNLP, Argentina
\and
Instituto de Astronomia, Geof{\'\i}sica e Ci\^encias Atmosf\'ericas, USP, Brasil
\and
Observatorio Nacional, Brasil
\and
Community Science and Data Center/NSF’s NOIRLab, Estados Unidos
}

\resumen{En el marco del Proyecto Fornax de la colaboraci\'on S-PLUS (S+FP por sus siglas en ingl\'es), analizamos la población de galaxias en la dirección del cúmulo de Fornax ($D\approx 20$~Mpc).
Contamos con 23 campos de $1.4^{\circ}\times 1.4^{\circ}$ que cubren la posición espacial proyectada de 999 galaxias de Fornax previamente reportados en la literatura. 244 de estas galaxias fueron identificadas en los campos de S-PLUS con fotometría confiable en las 12 bandas \'opticas del relevamiento. 
Obtuvimos, adem\'as, los parámetros estructurales y fotom\'etricos de $\approx 3\times10^5$ galaxias detectadas en dichos campos. En este trabajo presentamos resultados preliminares de la caracterización de la población de galaxias del cúmulo de Fornax con respecto a la población de galaxias de fondo. Entre otros objetivos, esperamos que dicha caracterización provea criterios fotométricos para identificar nuevos candidatos a ser galaxias miembro del cúmulo.
}

\abstract{In the context of the S-PLUS Fornax Project (S+FP), we analyze the galaxy population in the direction of the Fornax cluster ($D\approx 20$~Mpc). We have 23 fields of size $1.4^{\circ}\times 1.4^{\circ}$, covering the projected positions of 999 Fornax galaxies reported in the literature. 244 of those galaxies are detected with confident photometry in our fields which were observed simultaneously in 12 photometric bands. 
Besides those of Fornax galaxies, we obtained confident structural and photometric parameters for $\approx 3\times10^5$ additional galaxies detected in our fields. 
In this work we present preliminary results on the characterization of the galaxy population of the Fornax cluster with respect to the background galaxy population. Among other goals, we expect that such a characterization provides photometric criteria to identify new candidate members of the cluster.
}


\begin{document}

\maketitle

\section{Introduction}\label{sec:introduction}
The Southern Photometric Local Universe Survey (S-PLUS) aims at mapping 9300 squared degrees of the southern sky using 12 photometric bands in the optical range: 5 broad bands similar to those used by the Sloan Digital Sky Survey (SDSS; ugriz) and 7 narrow bands tracing specific spectral features ([OII], Ca H+K, H$\delta$, G-band, Mgb triplet, H$\alpha$ and Ca triplet). That goal will be achieved performing 5 sub-surveys: the Main Survey (MS), the Ultra-Short Survey (USS), the Galactic Survey (GS), the Marble Field Survey (MFS) and the Variability Fields Survey (VFS). For details about S-PLUS and its sub-surveys, we refer the reader to \citet{Oliveira2019}.

In the context of the MS, 23 S-PLUS fields were observed in the region of the Fornax cluster and a project aimed at performing the analysis of Fornax using those data (the S-PLUS Fornax Project; hereafter S+FP) started in 2020 \citep{SmithCastelli2021}. In this work we present preliminary results on the analysis of photometric and structural parameters of a sample of galaxies in the direction of the Fornax cluster performed in the context of the S+FP. 
As typical photo-z errors obtained using S-PLUS data are 2-3 times greater than the Fornax mean redshift ($\sigma_z=0.02-0.03$, $\langle z\rangle_{\rm Fornax} \sim 0.005$; \citealt{2022A&C....3800510L}),
it is not possible to identify new Fornax members from photo-zs. However, characterizing the galaxy population of Fornax from the background using the 12-bands photometry of S-PLUS, might allow us to establish photometric criteria to identify new Fornax members.
Through out this contribution we will consider $(m-M)=31.51$ for Fornax \citep{Blakeslee2009} and we will assume $H_{0} = 70.5$\,km\,s$^{-1}$\,Mpc$^{-1}$ \citep{2009ApJS..180..330K}.

\section{Data}

\begin{figure}
    \centering
    \includegraphics[width=0.45\textwidth]{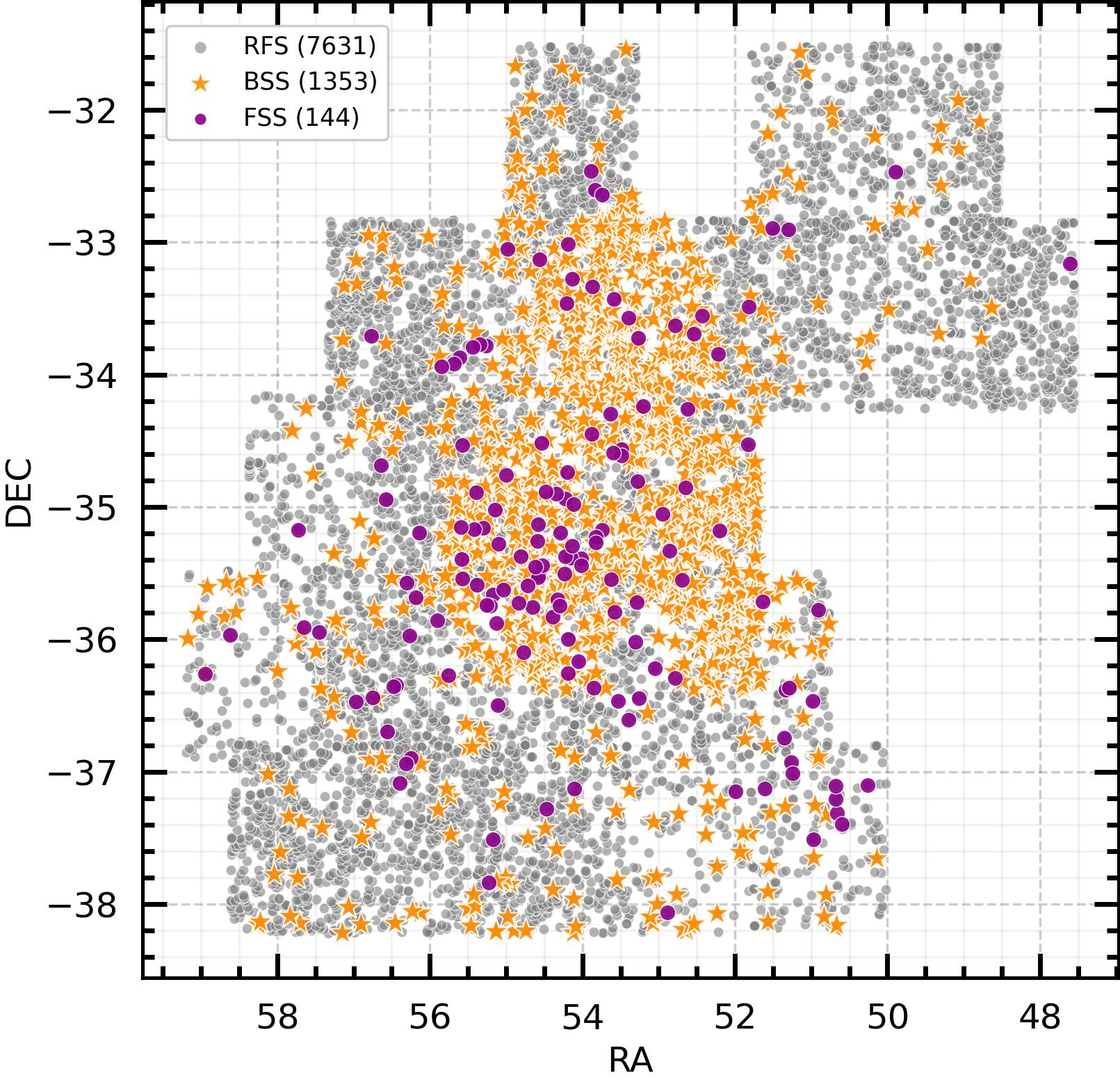}
        \caption{Spatial distribution of the restricted full sample (RFS; grey), the background spectroscopic sample (BSS; orange) and the Fornax spectroscopic sample (FSS; violet). These are the three galaxy samples used in our analysis.} 
        \label{fig:position}
\end{figure}

As part of the S+FP, we have calibrated photometry and structural parameters for $\sim663000$ objects detected in the 23 S-PLUS Fornax fields using {\sc SExtractor} \citep{Bertin1996}.
We further refer to this sample as the full sample. In addition, we compiled 1057 galaxies reported in the literature as Fornax members or candidates among which 999 are placed in our fields. The rest of the objects (58 galaxies) are in sky regions not yet observed by S-PLUS. All these galaxies were visually inspected one by one in the aperture images obtained from {\sc SExtractor} in order to check their correct detection in our images (that is, that {\sc SExtractor} apertures are well centered in the objects and properly cover the extension of the objects). Only those galaxies with reliable apertures and, as a consequence, with reliable S-PLUS photometry, will be considered in our study. From our visual inspection, we obtained a galaxy sample of 244 galaxies reported in the literature with confident S-PLUS photometry and we will refer to those galaxies as our Fornax cluster sample (FCS). 144 of those galaxies are spectroscopically confirmed Fornax members ($600~ \mathrm{km\,s}^{-1} < cz < 3000~\mathrm{km\,s}^{-1}$, \citealt{Maddox2019}) and we will refer to this spectroscopic sample as our Fornax spectroscopic sample (FSS).

\begin{figure}
    \centering
    \includegraphics[width=0.435\textwidth]{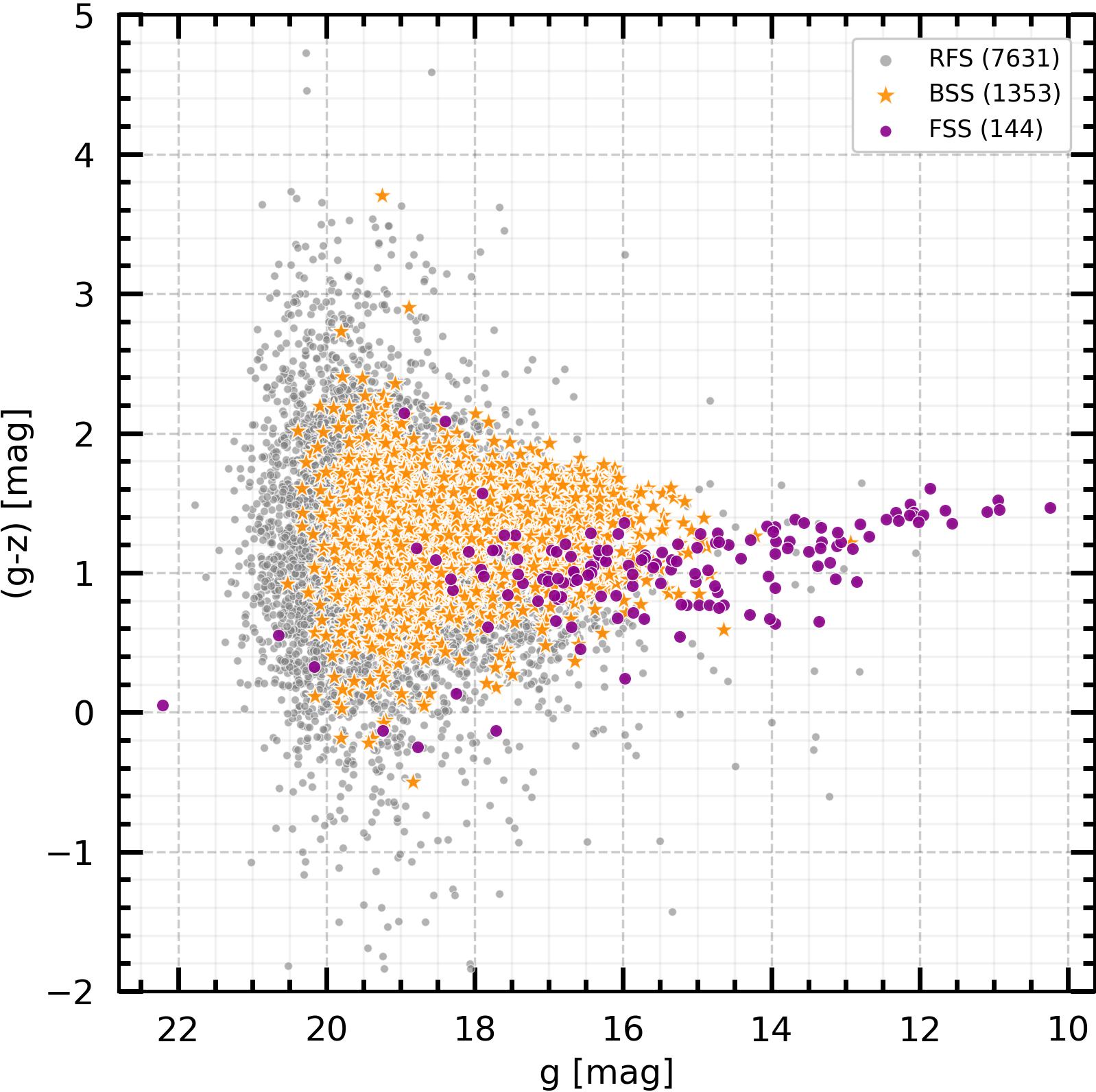}
    \caption{Colour-magnitude relation of the RFS, the BSS and the FSS. It can be seen that spectroscopic confirmed members of the Fornax cluster define a tight color-magnitude relation in comparison with spectroscopically confirmed background galaxies. The color code is the same as in Figure\,\ref{fig:position}}
    \label{fig:CMR}
\end{figure}

In order to discriminate unresolved sources from extended ones in the full sample, we used $\mathrm{CLASS\_STAR}<0.5$ from {\sc SExtractor} in the S-PLUS gri bands. The calibrated photometric errors of the FCS were also used to filter the full sample to discard spurious detections. From these criteria, we obtained a restricted full sample (RFS) of 7631 objects. Among them, 1353 are spectroscopically confirmed background galaxies and we will refer to this sub-sample as our background spectroscopic sample (BSS). In Figure\,\ref{fig:position}, we show the projected spatial distribution of the RFS (grey), the BSS (orange) and the FSS (violet) which are the three galaxy samples analyzed in this work, while in Figure\,\ref{fig:CMR} we show the location of the three samples in the color-magnitude diagram (CMD). From the CMD it can be seen that the FSS define a tight sequence in comparison to the distribution of the BSS.
\section{Results}
It is well known that the color-magnitude relation of galaxy clusters can be used as a membership criteria to identify new cluster members (see, for example, \citealt{Chiboucas2010} for the Coma cluster). Using the 12-bands photometry provided by S-PLUS, we aim at characterizing the galaxy population of the Fornax cluster from the background in order to establish additional photometric criteria to identify new Fornax members. To achieve our goal, we plan to use all the geometric and photometric parameters measured by {\sc SExtractor}. Among others, they include magnitudes and effective radius ($r_e$\footnote{FLUX\_RADIUS\_2 in {\sc SExtractor} which is the radius enclosing 50\% of the light of an object.}) that will be used simultaneously to build membership criteria avoiding selection effects.

\begin{figure*}
    \centering
    \includegraphics[width=0.44\textwidth]{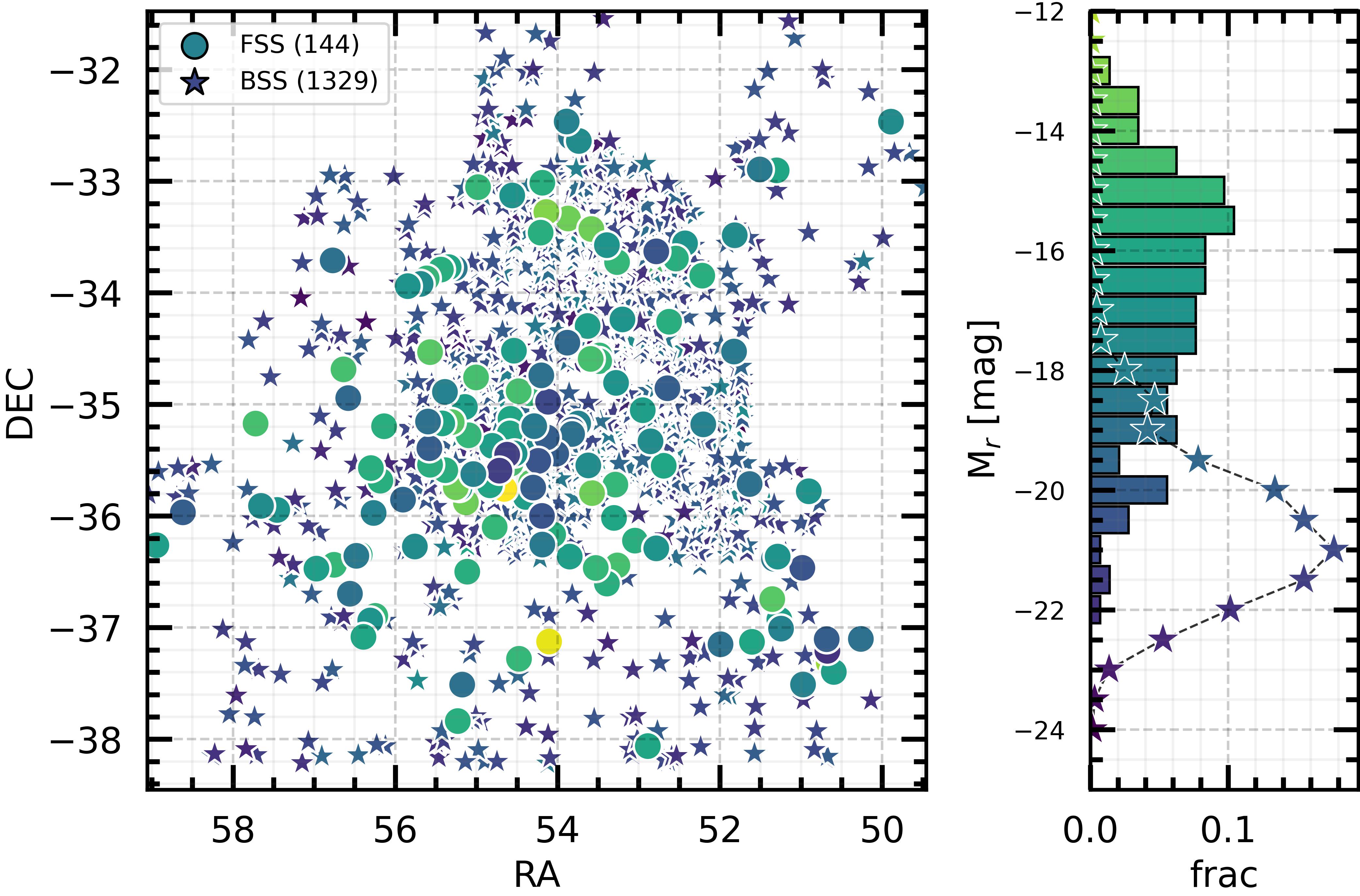}\qquad
        \includegraphics[width=0.44\textwidth]{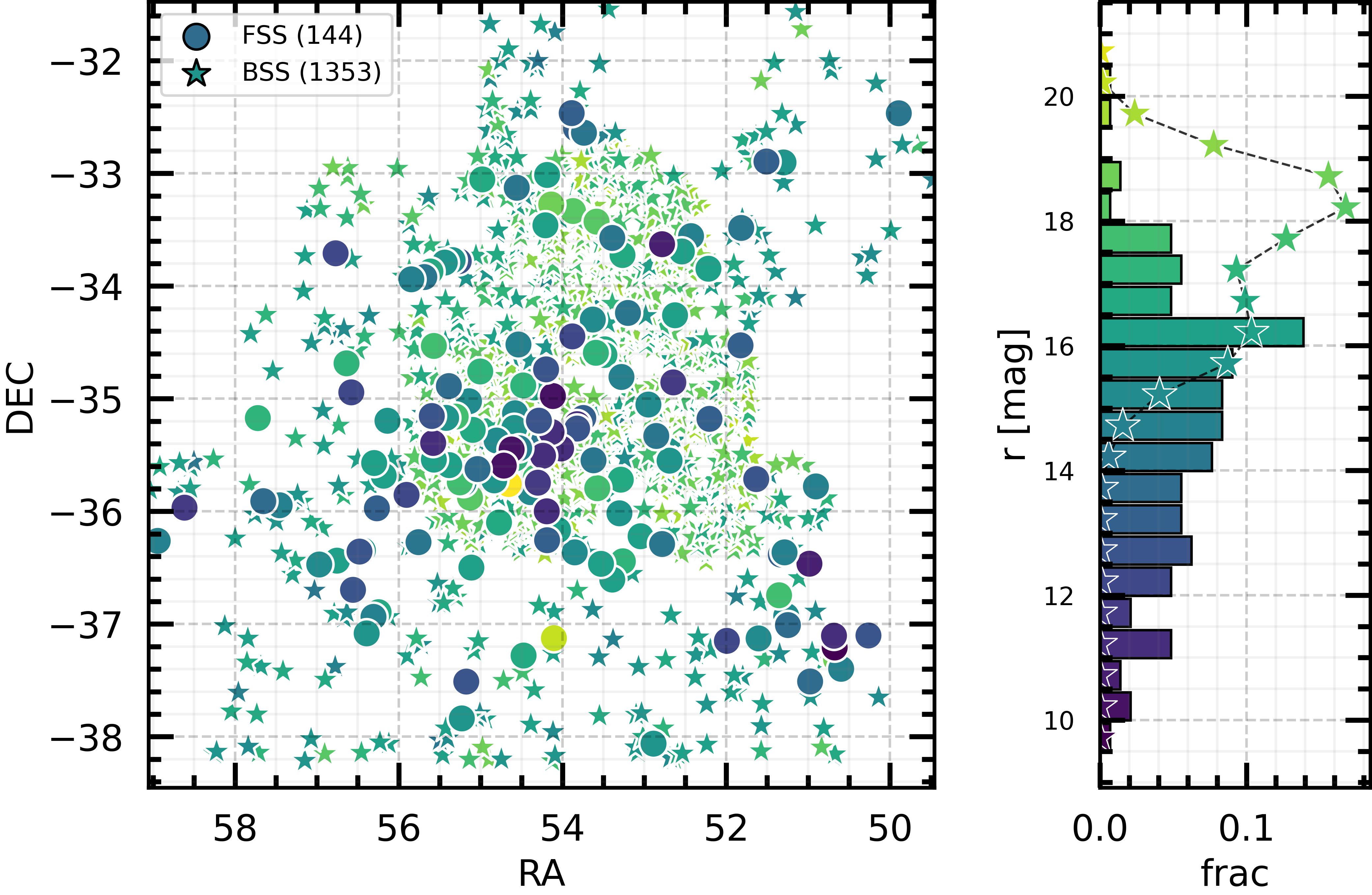}\\
    \includegraphics[width=0.44\textwidth]{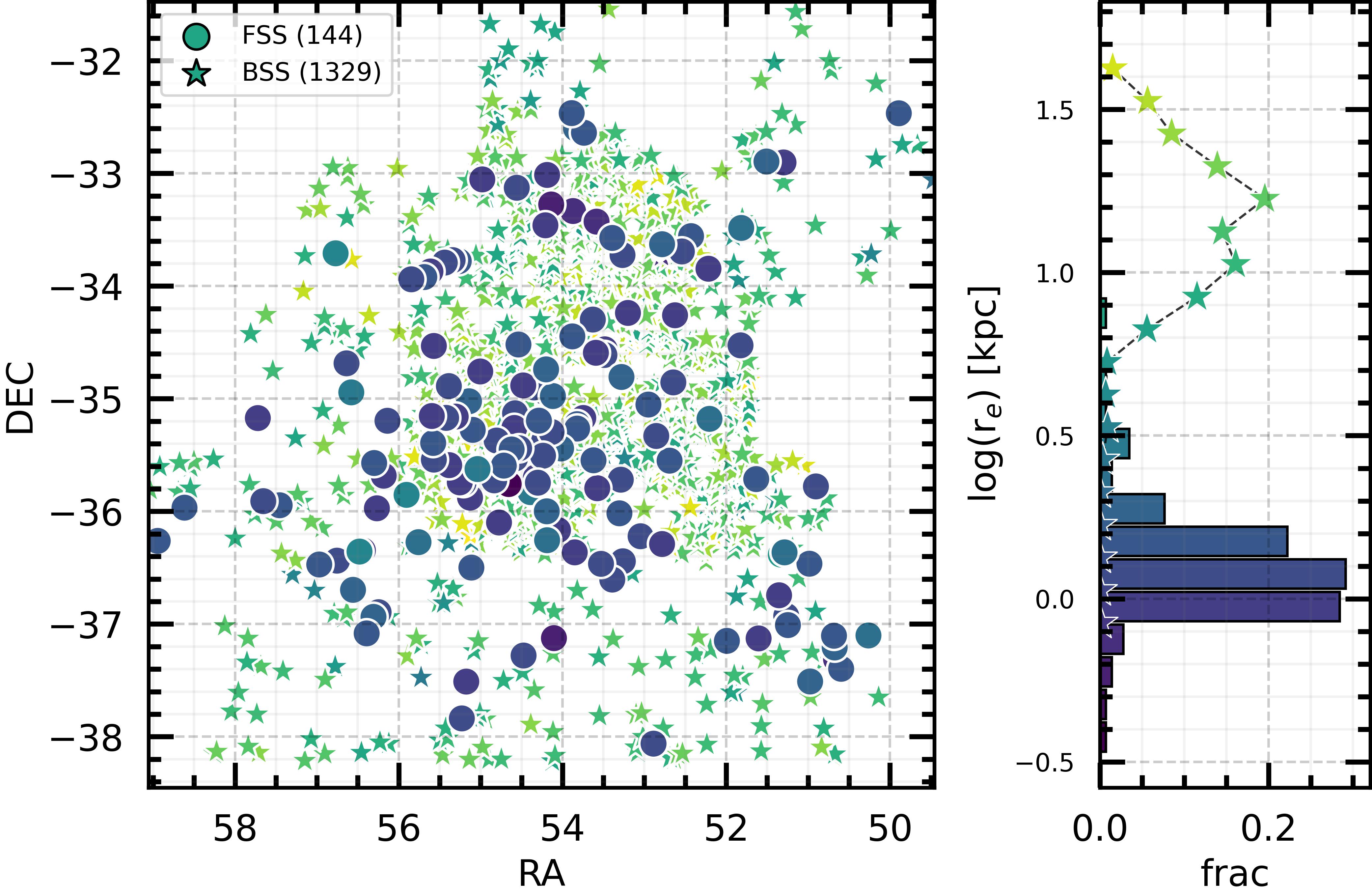}\qquad
        \includegraphics[width=0.44\textwidth]{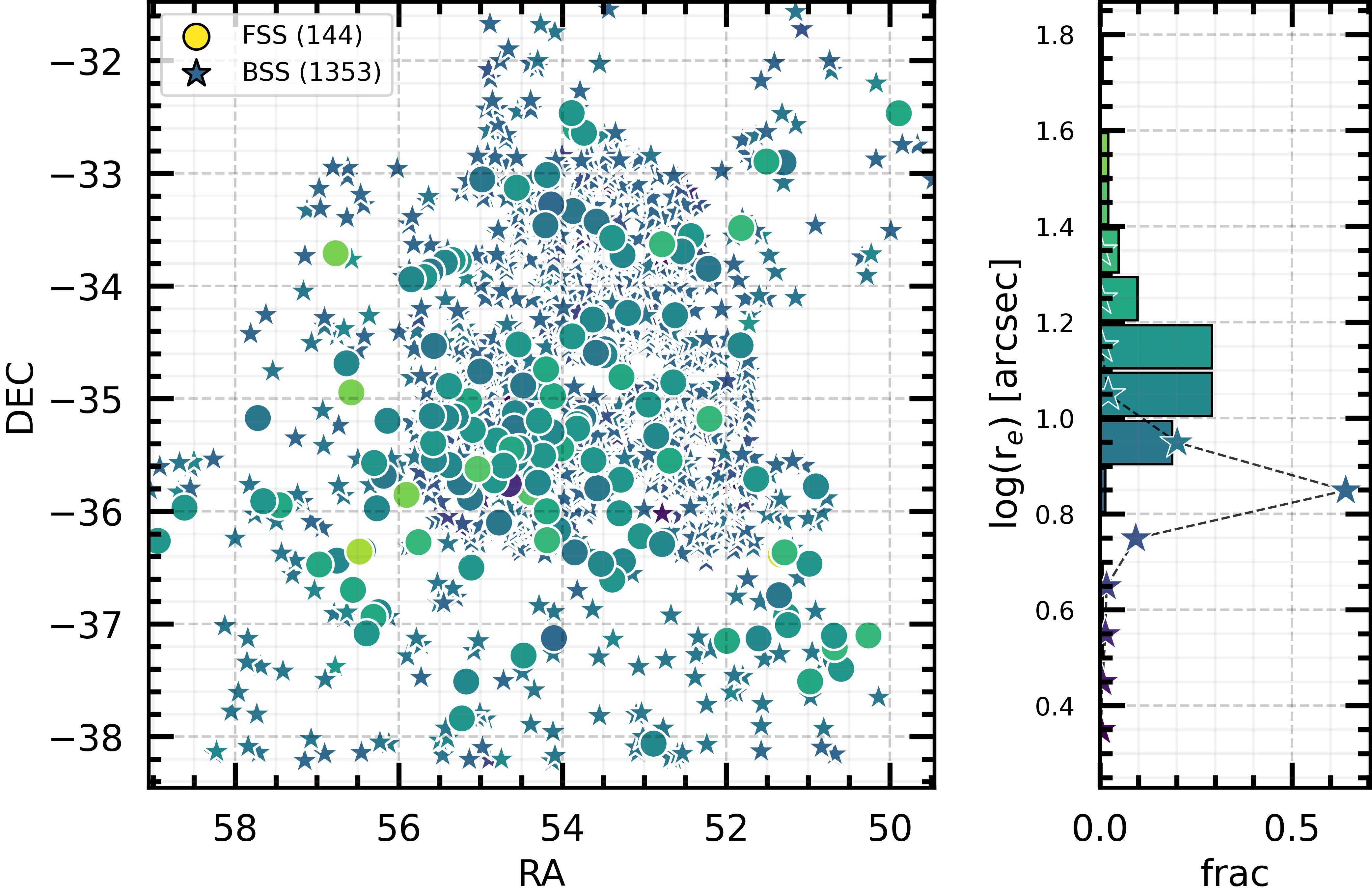}\\
    \caption{{\it Top panels:} Spatial projected distribution of the FSS and BSS colored according to the absolute ({\it left}) and the apparent ({\it right}) total ({\sc SExtractor} auto) magnitudes in the $r$-band (see the histograms at the left of each panel). The big circles in the spatial distribution and the colored bars in the histograms correspond to the FSS. The BSS is shown as stars in both plots. {\it Bottom panels:} The same as the top panels but for absolute and apparent effective radius ($r_e$).}
    \label{fig:parameters}
\end{figure*}

In Figure\,\ref{fig:parameters} we present four panels showing separately the spatial projected distribution of the FSS and BSS, and the histograms where their similitudes and differences can be seen. The big circles on the left of each panel, and the bars in the histograms, correspond to the FSS, while the stars on both plots correspond to the BSS. The top panels display the distribution of absolute ({\it left}) and apparent ({\it right}) r-band magnitudes while the bottom panels correspond to absolute ({\it left}) and apparent ({\it right}) $r_e$ in the r-band.

We can see that both samples are well separated in all those quantities. While the absolute parameters will allow us to characterize Fornax with respect to the background population, the apparent information will allow us to identify the values ranges expected for new Fornax candidates. As it can be seen from the top left panel of Figure\,\ref{fig:parameters}, the FSS displays a broad and quite homogeneous distribution in the range  $-22.5~\mathrm{mag}\lesssim Mr \lesssim-13.5~ \mathrm{mag}$, with a peak at $Mr=-15~\mathrm{mag}$, which corresponds to a typical brightness of dwarf galaxies. On the contrary, the BSS shows a tighter luminosity distribution in which most of the galaxies displays $-24~\mathrm{mag}\lesssim Mr\lesssim-17~\mathrm{mag}$, with a clear peak at $Mr\approx-21~\mathrm{mag}$, indicating that this population is dominated by luminous galaxies, as expected. 

In the case of the apparent r-band luminosity (top right panel of Figure\,\ref{fig:parameters}), the FSS shows a clear peak at $r \approx 16.5$\,mag, while the BSS have a peak at a fainter magnitude $r \approx 18.5$\,mag. In a search of new Fornax members, objects with $r>18~\mathrm{mag}$ should be considered with caution as galaxies with those brightnesses and radial velocities all lie in the background. However, other parameters should be analyzed before rejecting those objects as likely Fornax members.

In that sense, $r_e$ could be a helpful parameter. From the bottom left panel of Figure\,\ref{fig:parameters}, we can see that the FSS displays a tight distribution around a peak of $log(r_e)\approx0.1$ or $r_e\approx1.2~\mathrm{kpc}$, a behaviour that it was observed in other galaxy clusters like Virgo \citep{SmithCastelli2013}. On the contrary, the BSS is dominated by larger galaxies with a broader distribution where most of the objects display  $0.8\lesssim log(r_e)\lesssim1.6$ or $6.3~\mathrm{kpc}\lesssim r_e\lesssim39.8~\mathrm{kpc}$. In apparent sizes, Fornax is located in the range $1.0\lesssim log(r_e)\lesssim1.8$ or $10~\mathrm {arcsec}\lesssim r_e \lesssim63~\mathrm{arcsec}$ while the background depicts mainly $0.6\lesssim log(r_e)\lesssim1.2$ or $4~\mathrm{arcsec}\lesssim r_e \lesssim16~\mathrm{arcsec}$. Therefore, from the plots presented in Figure\,\ref{fig:parameters}, we could say that, at first order, bona-fide Fornax candidate members should be searched within the objects displaying $9~\mathrm{mag} \lesssim r \lesssim 18~\mathrm{mag}$ and $10~ \mathrm{arcsec}\lesssim r_e \lesssim63~\mathrm{arcsec}$.
\begin{acknowledgement}
We thank the anonymous referee for helping us to improve this work. S-PLUS is an international collaboration founded by Universidade de Sao Paulo, Observat\'orio Nacional, Universidade Federal de Sergipe, Universidad de La Serena and Universidade Federal de Santa Catarina. This work
was funded with grants from Consejo Nacional de Investigaciones
Cient\'ificas y T\'ecnicas, Agencia Nacional de Promoci\'on de la Investigaci\'on, el Desarrollo Tecnol\'ogico y la Innovaci\'on and Universidad Nacional de La Plata (Argentina).
\end{acknowledgement}
\bibliographystyle{baaa}
\small
\bibliography{bibliografia}
\end{document}